\documentclass[letterpaper]{article} 
\usepackage{aaai25}  
\usepackage{times}  
\usepackage{helvet}  
\usepackage{courier}  
\usepackage[hyphens]{url}  
\usepackage{graphicx} 
\usepackage{natbib}  
\usepackage{caption} 
\usepackage{algorithm}
\usepackage{algorithmic}
\usepackage{cite}
\usepackage{amsmath}
\usepackage{amssymb}
\usepackage{multirow}
\usepackage{subcaption}
\usepackage{newfloat}
\usepackage{listings}
\DeclareCaptionStyle{ruled}{labelfont=normalfont,labelsep=colon,strut=off} 
\floatstyle{ruled}
\newfloat{listing}{tb}{lst}{}
\floatname{listing}{Listing}
\title{Backdoor Token Unlearning: Exposing and Defending Backdoors in Pretrained Language Models}

\author{
    Peihai Jiang\textsuperscript{\rm 1}, Xixiang Lyu\textsuperscript{\rm 1}\thanks{Corresponding authors}, Yige Li\textsuperscript{\rm 2}\footnotemark[1], Jing Ma\textsuperscript{\rm 1}
}
\affiliations{
    \textsuperscript{\rm 1}Xidian University, China\\
    \textsuperscript{\rm 2}Singapore Management University, Singapore\\
    \{xdjph, jing.ma\}@stu.xidian.edu.cn, xxlv@mail.xidian.edu.cn,  yigeli@smu.edu.sg

}

\usepackage{bibentry}

\begin{document}

\maketitle

\begin{abstract}
Supervised fine-tuning has become the predominant method for adapting large pretrained models to downstream tasks. However, recent studies have revealed that these models are vulnerable to backdoor attacks, where even a small number of malicious samples can successfully embed backdoor triggers into the model. While most existing defense methods focus on post-training backdoor defense, efficiently defending against backdoor attacks during training phase remains largely unexplored. To address this gap, we propose a novel defense method called \textit{Backdoor Token Unlearning (BTU)}, which proactively detects and neutralizes trigger tokens during the training stage. Our work is based on two key findings: 1) backdoor learning causes distinctive differences between backdoor token parameters and clean token parameters in word embedding layers, and 2) the success of backdoor attacks heavily depends on backdoor token parameters. The BTU defense leverages these properties to identify aberrant embedding parameters and subsequently removes backdoor behaviors using a fine-grained unlearning technique. Extensive evaluations across three datasets and four types of backdoor attacks demonstrate that BTU effectively defends against these threats while preserving the model’s performance on primary tasks. Our code is available at \url{https://github.com/XDJPH/BTU}.

\end{abstract}

%

\section{Introduction}
Pretrained Language Models (PLMs)~\cite{BERT,GPT2} have demonstrated remarkable performance across various tasks, such as sentiment analysis~\cite{sentiment}, toxicity detection~\cite{Toxic}, and news classification~\cite{newsclassification}.  However, as PLMs are increasingly fine-tuned for specific downstream applications~\cite{survey}, they have become vulnerable to backdoor attacks ~\cite{BackdoorSurvey,backdoorattacksurvey}. Typically, backdoor attacks inject malicious triggers into the model during training. The backdoored model functions normally on clean tasks but exhibits an attack-desired target label when the trigger is presented. In Natural Language Processing (NLP), backdoor triggers can be designed as obvious elements like rare words~\cite{word} or more subtle features such as sentence styles~\cite{Stylebkd}. With the widespread adoption and deployment of PLMs, defending against backdoor threats has become an urgent challenge.

Existing backdoor defense methods in NLP generally fall into three categories: backdoor detection~\cite{backdoordetection1,PICCOLO,inferencedefense1}, backdoor removal~\cite{finemixing, NAD}, and anti-backdoor learning~\cite{ABL,MF}. Backdoor model detection methods aim to identify whether a model or inputs contain backdoors, while backdoor removal methods focus on purifying the backdoor triggers from the backdoored model. Among them, anti-backdoor learning methods~\cite{MF, ABL} has become a widely adopted defense strategy as they allow the users to train a clean model even on a poisoned dataset. For example, ABL~\cite{ABL} employs a two-stage gradient ascent technique to filter out and mitigate backdoor behaviors. Another approach, MF~\cite{MF}, limits the model's learning capacity by restricting the number of training epochs, thereby preventing the model from acquiring backdoors during training. However, these anti-backdoor learning methods often lead to reduced model performance and exhibit instability across different scenarios. Therefore, how to effectively defend against backdoor attacks during the model training phase essentially deserves much attention.

Previous research has shown that backdoor learning can be viewed as a dual-task problem, i.e. training the backdoored model on both clean and backdoor data~\cite{li2023reconstructive}. In this paper, we reformulate backdoor learning from model parameter perspective and identify two key properties: 1) backdoor learning induces significant differences between the embedding parameters of backdoor tokens and clean tokens, where the backdoor tokens converge much faster than clean ones; 2) the activation of backdoors is highly dependent on backdoor token parameters in the embedding layers. Intuitively, if we can isolate backdoor token parameters at the level of word embedding dimensions rather than across all model parameters, the backdoor information could be more effectively exposed and removed.

In this work, we propose a novel defense method called Backdoor Token Unlearning (BTU) for efficient anti-backdoor learning. Specifically, BTU operates in two stages: \textit{backdoor token detection} and \textit{dimensional fine-grained unlearning}. In the first stage, BTU identifies potential backdoor tokens by exclusively training the word embedding layer and flagging the top $\alpha$\% as backdoor-related embedding parameters. In the second stage, BTU removes backdoor information by replacing the affected backdoor embedding parameters with those of benign padding token embeddings. Through these two stages, BTU effectively defends against backdoor attacks while minimizing the impact on clean task performance. The main contributions of our work are summarized as follows:

\begin{itemize}
\item We identify two key observations in NLP backdoor attacks: 1) the distinctive differences in the embedding values of backdoor tokens and clean tokens when only the word embedding layer is trained, and 2) the success of backdoor activation is highly related to the backdoor token embedding parameters.

\item We introduce a novel defense method termed Backdoor Token Unlearning (BTU), which proactively exposes aberrant embedding parameters of backdoor tokens and mitigates backdoor behavior during the training process, with minimal impact on clean task performance.

\item Extensive experiments on four types of backdoor attacks across three datasets demonstrate that our proposed BTU substantially reduces the success rate of backdoor attacks while having minimal impact on the accuracy of downstream tasks.
\end{itemize}

\section{Related Work}
\subsection{Backdoor Attack}
Existing backdoor attacks in NLP manifest in two primary scenarios: outsourced training and data poisoning. In outsourced training, attackers have full control over the training process.
For instance, the LWP~\cite{layerwise} scheme implants backdoors in the model's intermediate layers to increase the persistence of the attack, while the transfer~\cite{transfer} approach adjusts the backdoor optimization target in front of the MLP layer, using a multi-objective strategy to ensure the attack's resilience against downstream task influences. Additionally, LWS~\cite{Combination} employs an auxiliary model to create more concealed triggers. Conversely, in data poisoning scenarios, attackers are limited to inserting a few carefully crafted samples into the dataset since they do not control the training process. For example, Dai et al.~\cite{sent} demonstrate that words or context-independent phrases can serve as triggers, and that random insertion into training samples can successfully inject backdoors. Similarity, Qi et al.~\cite{Stylebkd, Synbkd} reveal that textual styles and syntactic structures can also act as triggers, significantly enhancing the stealthiness of backdoor attacks. These studies highlight the high vulnerability of NLP models to such covert manipulations and underscore the critical need for robust defense mechanisms.

\subsection{Backdoor Defense}
In the field of NLP, existing backdoor defense methods can be broadly categorized into three types:
1) Backdoor input detection, which is applied during the model inference stage to identify and prevent the activation of backdoor inputs~\cite{Strip,BKI,RAP}. For example, BKI~\cite{BKI} distinguishes potential trigger words by analyzing each word's impact on the model's outcomes;
2) Backdoored model detection, which assesses whether a model contains backdoors~\cite{PICCOLO,T-miner}, often employing techniques like reverse engineering. For instance, PICCOLO attempts to recover potential triggers embedded within the model;
3) Anti-backdoor learning aims to train clean models from potentially poisoned datasets during the training phase~\cite{ABL,MF,survey}. For instance, ABL~\cite{ABL} characterizes backdoor learning as a form of shortcut learning, where backdoor triggers are more easily captured. To address this, ABL proposed a two-stage gradient ascent technique to mitigate backdoor effects. Similarly, the MF defense~\cite{MF} introduced to minimize overfitting to prevent the model from learning backdoor patterns. Although promising, these methods often fail against adaptive attacks, such as textual style or grammatical structure triggers. In this work, we present new insights into backdoor learning and propose an simple yet efficient anti-backdoor defense to mitigate such threat.

\begin{figure*}[t]
    \centering
    \begin{subfigure}[b]{0.9\textwidth}
        \includegraphics[width=\textwidth]{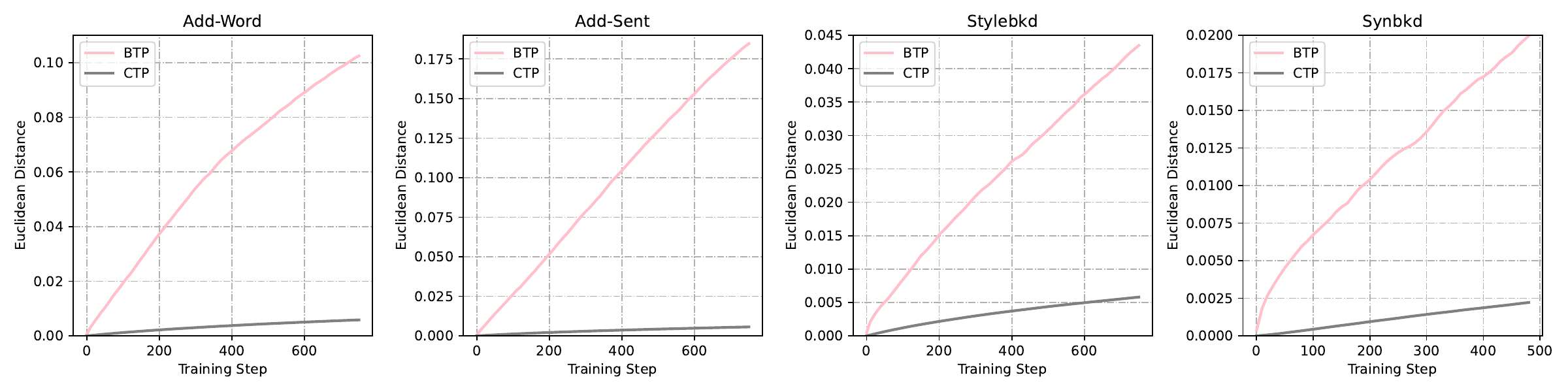}
        \caption{\textbf{Parameter: Embedding Layers}}
        \label{embedding}
    \end{subfigure}
    \\

    \begin{subfigure}[b]{0.9\textwidth}
        \includegraphics[width=\textwidth]{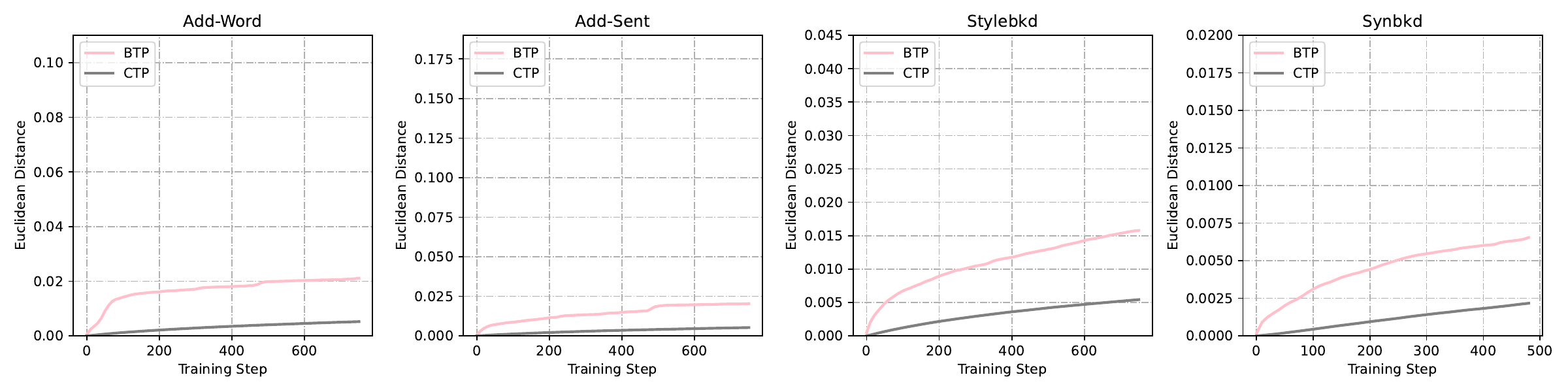}
        \caption{\textbf{Parameter: All layer}}
        \label{all}
    \end{subfigure}
    \caption{The distinctive learning behaviors for BTP and CTP under four different backdoor attacks. Figure (a) represents the variation in BTP and CTP as a function of the number of iterations when only optimizing the model's word embedding layer. Figure (b) represents the variation in BTP and CTP as a function of the number of iterations when optimizing all model parameters. In the Stylebkd and Synbkd attacks, conjunctions and punctuation marks are chosen as backdoor tokens. These abnormal behaviors are consistent across other attacks as well, highlighting the generalization of this phenomenon.
}
    \label{revealing}
\end{figure*}

\section{Proposed Token Unlearning Method}
In this section, we first present the problem of backdoor attacks and then reveal the distinctive behavior between backdoor tokens and clean tokens optimized in the word embedding layers. Finally, we introduce our proposed BTU method.
\subsubsection{Problem definition} 

Consider the poisoned training dataset as $\mathcal{D} = \mathcal{D}_c \cup \mathcal{D}_b$, where $\mathcal{D}_c$ denotes the subset of clean data and $\mathcal{D}_b$ denotes the subset of backdoor data. Training a backdoored model on a poisoned dataset can be viewed as minimizing the following empirical error:

\begin{equation}
\begin{gathered}
\mathcal L = 
 \underbrace {{\mathbb{E}_{(x,y) \sim \mathcal{D}_c}}[\ell({f_\theta }(x),y)]}_{\text{clean task}} + \underbrace {{\mathbb{E}_{(x,y) \sim \mathcal{D}_b}}[\ell({f_\theta }(x),y)]}_{\text{backdoor task}},
\end{gathered}
\label{eq:backdoor_learning}
\end{equation}

\noindent where $\ell$ and $\theta$ denote the loss function and model parameters, respectively. The overall learning task can be regarded as a combination of the backdoor task on dataset $\mathcal{D}_b$ and the clean task on dataset $\mathcal{D}_c$. 

Intuitively, if we can clearly distinguish between clean and backdoor tasks, the backdoor task can be more effectively detected. To achieve this, we reformulate the backdoor learning process in Eq.~\ref{eq:backdoor_learning} to focus on the word embedding layer rather than all model parameters. As a result, the model's optimization objective can be redefined as follows:

\begin{equation}
\begin{gathered}
\mathcal L = 
 \underbrace {{\mathbb{E}_{(x,y) \sim \mathcal{D}_c}}[\ell(\boldsymbol{\varepsilon }(x),y)]}_{\text{clean task}} + \underbrace {{\mathbb{E}_{(x,y) \sim \mathcal{D}_b}}[\ell(\boldsymbol{\varepsilon^b}(x),y)]}_{\text{backdoor task}},
\end{gathered}
\label{eq:embedding_learning}
\end{equation}

\noindent where $\varepsilon$ denotes the entire clean embedding parameters and $\varepsilon^b$ denotes backdoor embedding parameters.
Based on Eq.~\ref{eq:embedding_learning},  the backdoor information is primarily contained in the \textit{Backdoor Token Parameters (BTP)}, while the \textit{Clean Token Parameters (CTP)} remain largely unchanged.  Since the backdoor task is much simpler than the clean task~\cite{ABL}, we observe that the cumulative parameter changes in BTP occur more rapidly than in CTP. We will provide empirical evidence to support this observation in the following subsection.

\subsection{Revealing Distinctive Behavior of Backdoor Tokens} 
\label{sec:distinct_behavior}

In this subsection, we aim to highlight the distinct learning behavior between BTP and CTP when trained on word embedding layers.

We conduct four backdoor attack methods: Add-Word~\cite{badnets}, Add-Sent~\cite{sent}, Stylebkd~\cite{Stylebkd}, and Synbkd~\cite{Synbkd}, to poison the SST-2 dataset~\cite{SST-2} with a 10\% poisoning rate. We then train a BERT~\cite{BERT} model using standard procedures and settings from the public library~\cite{Openbackdoor}. For each attack, we trained two backdoored models: one on all parameters and another only on the word embedding layers. To compare the learning differences, we record the variations in Euclidean distance between BTP and CTP. 

Fig.~\ref{revealing} shows that, across all four types of attacks, the mean Euclidean distance of the BTP is greater than that in the CTP. For example, in Add-Word attack, when training only the word embedding layer, BTP is almost 0.1 higher than CTP. However, when training all parameters, BTP is only 0.01 higher than CTP. The difference in the magnitude of change between the two cases is nearly tenfold. This distinction between BTP and CTP suggests that backdoor information is primarily associated with BTPs and inspires our defense strategy.

\begin{figure*}[t]
\centering
\includegraphics[width=0.95\textwidth]{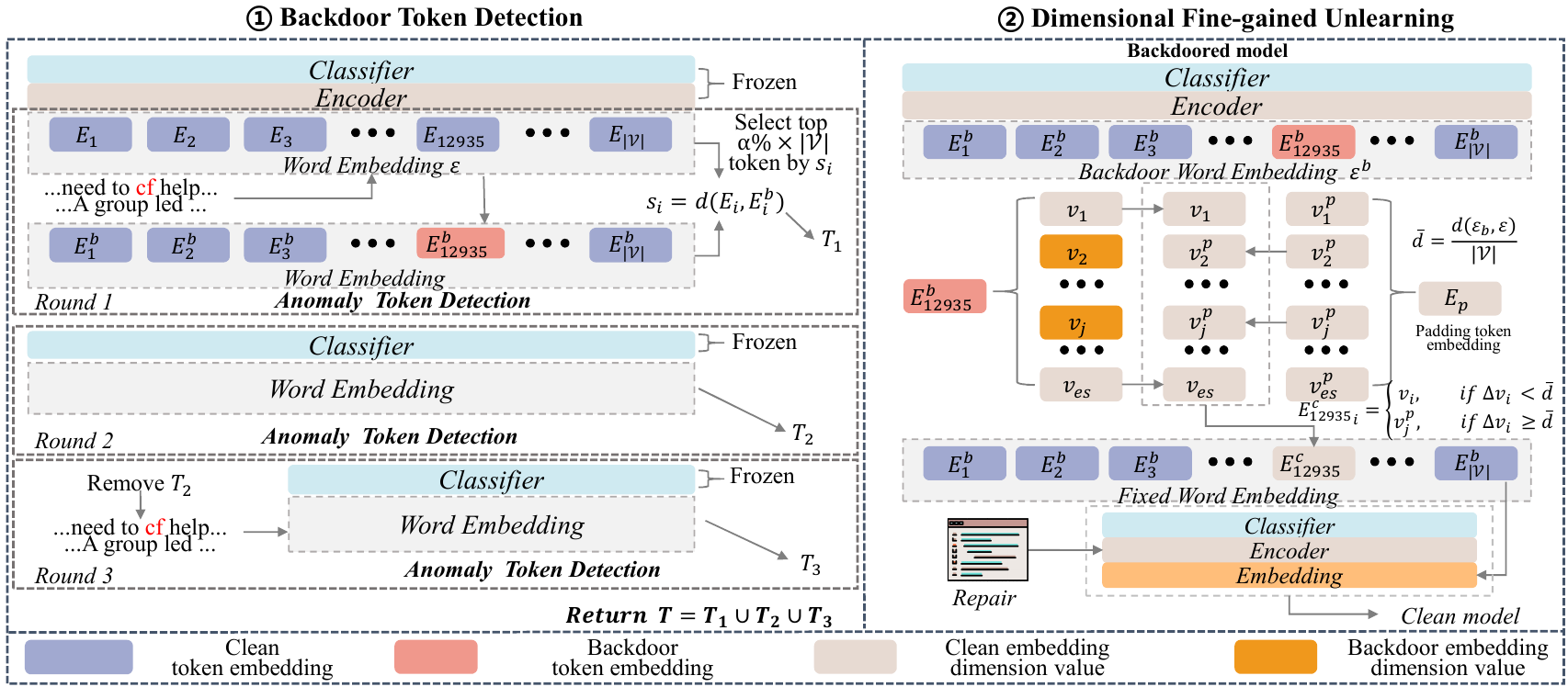} 
\caption{Illustration of the BTU Framework: 1) The Backdoor Token Detection phase includes three rounds of Anomaly Token Detection, where only the embedding layer is trained to detect embeddings with significant changes. 2) Dimensional Fine-grained Unlearning uses padding token embedding to precisely correct the anomalies tokens detected by backdoor token detection, and then the model is repaired using clean data.}

\label{BTU}
\end{figure*}

\subsection{Backdoor Token Unlearning}
\subsubsection{Overview} 
Fig.~\ref{BTU} illustrates the BTU framework, which consists of two main components: \textit{Backdoor Token Detection} and \textit{Dimensional Fine-grained Unlearning}. The backdoor token detection aims to identify suspicious backdoor tokens within the embedding parameters through three rounds of anomaly detection. Once these malicious tokens are detected, fine-grained dimensional unlearning is applied to remove backdoor functionalities from these token parameters. We provide detailed technical explanations below.

\subsubsection{Backdoor Token Detection} 

As previously noted, we have identified a distinctive Euclidean distance between BTP and CTP.  Building on this, we can detect suspicious backdoor token parameters through iterative detection rounds $T$. The detection threshold is set to $\alpha \in [0, 1]$, with the top $\alpha$\% of embedding parameters flagged as backdoor token parameters in each detection round. For simplicity, we set $\alpha$ to 0.05 across all three detection rounds. A more detailed analysis of $\alpha$ and the detection round $T$ will be provided in the ablation study. 

In the first round, we train only the embedding layer parameters $\varepsilon$ of model $M$ on the dataset $\mathcal{D}$, resulting in the updated embedding layer parameters $\varepsilon'$. We then calculate the change distance $s'$ for each token $t_i$ and store the token-distance pairs in the set $T_1$:
\begin{equation}
T_1 = \{(s_i', t_i) \}_{ t_i \in \mathcal{V}} = \{(d(\varepsilon(t_i), \varepsilon'(t_i)), t_i) \}_{ t_i \in \mathcal{V}}
\label{t1}
\end{equation}
Next, we rank $s_i'$ in descending order and select the top $\alpha\% \times |\mathcal{V}|$ tokens from $T_1$, which we denote as $T'$. 

In the second round, we retain the embedding layer and classification head of model $M$ , denoted as $M^*$, and train the embedding layer $\varepsilon$ of $M^*$ to obtain $\varepsilon''$. After training on the dataset $\mathcal{D}$, we calculate the change distance $s''$ for each token and store the token-distance pairs in the set $T_2$:
\begin{equation}
T_2 = \{(s_i'', t_i) \}_{ t_i \in \mathcal{V}} = \{(d(\varepsilon(t_i), \varepsilon''(t_i)), t_i) \}_{ t_i \in \mathcal{V}}
\label{t2}
\end{equation}
We then rank $s_i''$ in descending order and select the top $\alpha\% \times |\mathcal{V}|$ tokens from $T_2$, denoted as $T''$. 

In the third round, we repeat the previous procedure, but modify the dataset to $\mathcal{D}/T''$. All other settings remain the same, leading to:
\begin{equation}
T_3 = \{(s_i''', t_i) \}_{ t_i \in \mathcal{V}} = \{(d(\varepsilon(t_i), \varepsilon'''(t_i)), t_i) \}_{ t_i \in \mathcal{V}}
\label{t3}
\end{equation}
Finally, we rank $s_i'''$ in descending order and select the top $\alpha\% \times |\mathcal{V}|$ tokens from $T_3$, denoted as $T'''$.

We define $T = T' \cup T'' \cup T'''$ as the set of suspicious tokens. 
Notably, the three rounds of anomaly detection serve different purposes. Rounds 1 and 2 aim to detect simple triggers, while Round 3 refines the process to detect more complex triggers. This three-step iterative detection ensures comprehensive identification of suspicious backdoor tokens, effectively exposing both simple and complex triggers at varying levels of granularity. The analysis of results for different detection rounds can be found in the ablation study.



\subsubsection{Dimensional Fine-gained Unlearning}
Given a backdoored model $M^b$ and a set of suspicious tokens $T$, the most straightforward method is to replace all tokens in $T$ with padding tokens that carry no information, thereby removing all backdoor-related token parameters. However, simple replacement would eliminate both backdoor and clean features within the word embedding parameters, leading to a decrease in model accuracy. 

To maximally retain clean features in the word embedding parameters, we propose a \textit{Dimensional Fine-grained Unlearning} technique, which allows selectively replace only the dimensions with large changes in BTP while remaining others unchanged. Specifically, we first calculate the mean change in the word embedding layer before and after training:
\begin{equation}
\bar{d}=\sum\limits_{t_i \in \mathcal{V}} (d(\varepsilon(t_i),\varepsilon'(t_i)))/|\mathcal{V}|,
\label{eq6}
\end{equation}
where $\varepsilon$ represents the parameters of the word embedding before training, and $\varepsilon'$ represents the parameters after training. 

For all $t \in T$, the dimensions in $\varepsilon'(t)$ with values greater than $\bar{d}$ are replaced by the corresponding dimension values of $\varepsilon'(p)$, where $p$ denotes the padding token. Thus, the suspicious parameters in embedding layers $\varepsilon^c(t)$ are replaced by:
\begin{equation}
\varepsilon_i^c(t)=
\begin{cases} 
\varepsilon_i'(t), & \text{if } |\varepsilon_i'(t) - \varepsilon_i(t)| < \bar{d}; \\
\varepsilon_i'(p), & \text{if } |\varepsilon_i'(t) - \varepsilon_i(t)| \geq \bar{d}.
\end{cases}
\label{eq7}
\end{equation}
Finally, the values in $\varepsilon'(t)$ are replaced with $\varepsilon^c(t)$.
As we replace only a small number of tokens and the word embedding layer contains relatively little downstream information, the impact of our token unlearning causes minimal degradation in clean performance. To further mitigate the negative effect, we fine-tune the model with a small amount of clean data after padding token replacement.

\section{Experiment}
\subsection{Experimental Setting}
\noindent\textbf{Datasets and Models} We conducted experiments using three text classification datasets: 1) SST-2 (Stanford Sentiment Treebank-2)~\cite{SST-2}, a binary sentiment analysis dataset; 2) OLID (Offensive Language Identification Dataset)~\cite{OLID}, a binary toxicity detection dataset; and 3) AG News, a four-class news headline classification dataset. The victim model used is BERT-BASE-UNCASED, which consists of 12 layers with 30522 $\times$ 768 parameters in the word embedding layer.

\noindent\textbf{Attack Setups} Four data poisoning-based attack methods are employed: 1) Add-Word, using rare words as triggers (e.g., ``cf", ``tq", and ``bb"); 2) Add-Sent, using common phrases as triggers (e.g., ``I watched a 3D movie"); 3) Stylebkd, using text styles as triggers (e.g., ``Bible style"); and 4) Synbkd, using syntactic structures as triggers (e.g., ``(ROOT (S (SBAR) (, ) (NP) (VP) (.)))"). The poisoned samples for Stylebkd and Synbkd are generated using the public library from Cui et al.~\cite{Openbackdoor}.

\noindent\textbf{Defense Setups} We compared BTU with nine other methods, including six training-phase defenses (BKI~\cite{BKI}, MF~\cite{MF}, CUBE~\cite{survey}, TG~\cite{TG}, ST~\cite{ST}, and DPOE~\cite{DPOE}) and three inference-phase defenses (ONION, RAP, and Strip), which were adapted into training-phase defenses using the public library~\cite{Openbackdoor} under standard settings. For BTU, we removed special tokens from the results of backdoor token detection to refine the evaluation.

\noindent\textbf{Evaluation Metrics} Defense methods are evaluated using the metric ACC (Accuracy), which measures the model's ability to correctly classify clean data, and the metric ASR (Attack Success Rate), which measures the effectiveness of the backdoor attack in causing misclassification.

\subsection{Experimental Results}

\begin{table*}[t]
    \centering
    \caption{The attack success rate (ASR\%) and the accuracy (ACC\%) of our BTU and other 9 different defense methods against 4 backdoor attacks. None means without defense.}
    \label{main}
    \begin{tabular}{ccccccccccc}
    \hline
    \multirow{2}{*}{\textbf{Dataset}} & \multirow{2}{*}{\textbf{Defense}}  & \multicolumn{2}{c}{\textbf{Add-Word}} & \multicolumn{2}{c}{\textbf{Add-Sent}} & \multicolumn{2}{c}{\textbf{Stylebkd}} & \multicolumn{2}{c}{\textbf{Synbkd}} \\
                             &                          & \textbf{ACC}          & \textbf{ASR}           & \textbf{ACC}          & \textbf{ASR}           & \textbf{ACC}          & \textbf{ASR}          & \textbf{ACC}          & \textbf{ASR}          \\
    \hline
    \multirow{8}{*}{SST-2}   & None     & 91.05 & 100.00 & 91.10 & 100.00 & 90.37 & 54.72 & 90.72 & 90.46  \\
                             & ONION    & 87.08 & 21.18 & 86.78 & 71.38 & 84.32 & 60.15 & 85.27 & 91.33  \\
                             & RAP      & 91.82 & 100.00 & 90.88  & 99.89  & 87.34 & 56.80 & 87.70 & 94.74  \\
                             & STRIP    & 91.05 & 100.00 & 90.88 & 99.89 & 87.34 & 56.80 & 90.22 & 86.51  \\
                             & BKI      & 87.12 & 25.43 & 91.21 & 97.48 & 89.76 & 57.49 & 88.96 & 93.64  \\
                             & CUBE     & 87.70 & 15.68 & 88.14 & 30.81 & 90.88 & 20.50 & 90.94 & 28.18  \\
                             & MF       & 90.05 & 16.59 & 91.05 & 90.89 & 90.48  & 58.37  & 90.71 & 48.60 \\
                             & ST     & 90.35 & 19.03 & 90.73 & 22.55 & 89.01 & 19.03 & 86.26 & 43.71 \\
                             & TG     & 88.37 & 19.45 & 88.19 & 20.91 & 89.09 & 27.98 & 89.22 & 37.93 \\
                             & DPOE   & 88.30 & 19.63 & 90.33 & 50.54 & 89.01 & 17.37 & 89.89 & 36.99 \\
                             & \textbf{BTU (ours)} & \textbf{90.37} & \textbf{5.97} & \textbf{90.69} & \textbf{5.50} & \textbf{90.38} & \textbf{6.79} & \textbf{90.59} & \textbf{24.36}  \\
    \hline
    \multirow{8}{*}{OLID}    & None     & 79.51 & 100.00 & 79.68 & 100.00  & 76.03 & 52.33 & 79.67 & 97.61 \\
                             & ONION    & 78.23 & 10.46 & 77.55 & 100.00 & 66.59 & 71.69 & 72.91 & 97.86  \\
                             & RAP      & 79.51 & 100.00 & 62.06 & 0.11 & 76.04 & 52.33 & 77.42 & 97.45  \\
                             & STRIP    & 79.52 & 100.00 & 75.36 & 94.40 & 76.04 & 52.33 & 79.00 & 93.78  \\
                             & BKI      & 75.13 & 25.26 & 79.51 & 100.00 & 69.76 & 70.65 & 70.87 & 94.58  \\
                             & CUBE     & 77.47 & 18.66 & 80.01 & 16.26 & 77.02 & 25.71 & 79.81 & 21.07  \\
                             & MF       & 79.19 & 19.71 & 79.13 & 81.66 & 75.99 & 43.78 & 79.55 & 56.90 \\
                             & ST     & 77.68 & 22.70 & 79.13 & 21.79 & 78.02 & 33.27 & 79.43 & 42.71 \\
                             & TG     & 77.54 & 13.80 & 77.76 & 15.35 & 76.04 & 26.32 & 78.06 & 38.07 \\
                             & DPOE   & 76.78 & 98.83 & 55.61 & 95.51 & 50.07 & 50.02 & 50.10 & 48.07 \\
                             & \textbf{BTU (ours)} & \textbf{78.93} & \textbf{4.04} & \textbf{79.12} & \textbf{5.17}  & \textbf{79.32} & \textbf{5.33} & \textbf{80.24} & \textbf{12.77}   \\
    \hline
    \multirow{8}{*}{AG News} & None     & 94.47 & 100.00 & 94.46 & 100.00 & 93.39 & 73.72 & 94.02 & 100.00 \\
                             & ONION    & 92.91 & 2.05 & 93.02 & 77.63 & 90.39 & 76.91 & 93.11 & 96.11  \\
                             & RAP      & 94.26 & 84.82 & 94.46 & 100.00 & 93.11 & 67.58 & 93.49 & 79.98 \\
                             & STRIP    & 94.33 & 99.98 & 94.25 & 100.00 & 93.33 & 74.02 & 93.41 & 79.81 \\
                             & BKI      & 94.11 & 96.15 & 94.15 & 100.00 & 93.00 & 76.15 & 93.27 & 82.67  \\
                             & CUBE     & 87.04 & 3.97 & 88.14 & 2.71 & 91.98 & 2.53 & 90.46 & 3.87  \\
                             & MF       & 94.31 & 17.79 & 94.13 & 89.37 & 92.97 & 66.34 & 93.71 & 68.77 \\
                             & ST     & 93.94 & 20.10 & 93.56 & 18.37 & 93.07 & 33.74 & 93.47 & 45.47 \\
                             & TG     & 91.08 & 2.39  & 91.20 & 1.90  & 90.47 & 11.79 & 92.68 & 40.83 \\
                             & DPOE   & 94.84 & 1.63  & 93.99 & 5.26  & 93.15 & 15.45 & 93.89 & 55.48 \\
                             & \textbf{BTU (ours)} & \textbf{94.35} & \textbf{0.83} & \textbf{94.33} & \textbf{1.58} & \textbf{93.90} & \textbf{11.47} & \textbf{93.58} & \textbf{37.39}  \\      
    \hline
\end{tabular}
\end{table*}

As shown in Table~\ref{main}, BTU significantly reduces the success rate of four types of backdoor attacks across three datasets. Specifically, for insertion-based attacks (Add-Word and Add-Sent), BTU reduces the ASR to below 10\% across all three datasets. Additionally, it is observed that the more complex the dataset, the more effective BTU becomes. 
Across all datasets, we find that the ACC of the Add-Sent attack is higher than that of the Add-Word attack. This is because BTU detects more clean tokens in the Add-Word attack, resulting in the loss of more clean features. 

For unfixed type triggers in Stylebkd and Synbkd, BTU successfully mitigate the influence of backdoor attacks, demonstrating that these backdoor attack activations still depend on specific tokens. This phenomenon can also be observed in the poisoned samples, where conjunctions such as "when" and "if" are frequently involved. Additionally, we find that Stylebkd negatively affects the model's performance; however, BTU can effectively restore the damage caused by this attack.

To explore the generalizability of the BTU method, we conducted experiments on GPT2, RoBERTa and LLaMA2-7B~\cite{llama2} with SST-2 dasaset under 10\% poison ratio. As summarized in Table~\ref{table2}, BTU significantly lowers the ASR on all models while maintaining high ACC.

Overall, the results demonstrate that BTU is effective in defending against a variety of known backdoor attacks across different attack scenarios, with minimal impact on the model's performance on clean tasks. This consistent performances across datasets and attack types highlights BTU's potential as a reliable defense mechanism in real-world applications, where maintaining accuracy while ensuring security is paramount. More results of BTU Additional results pertaining to BTU are provided in Appendix.

\begin{table}[t]
\caption{The performance of BTU under different model architectures. The experiments are conducted on SST-2 dataset with 10\% poisoning rate.}
\label{table2}
\centering
\begin{tabular}{cccc}
\hline
\textbf{Model}  & \textbf{Attack}   & \textbf{ACC} & \textbf{ASR} \\ \hline
\multirow{2}{*}{RoBERTa} & Add-Sent &  94.01   &   19.71  \\
                        & Synbkd   &  91.04   &   23.55  \\ \hline
\multirow{2}{*}{GPT2}   & Add-Sent &  90.51   &   25.78  \\
                        & Synbkd   &  89.87   &   29.98  \\  \hline
\multirow{2}{*}{LLaMA2}   & Add-Sent &  91.76   &   5.17  \\
                        & Synbkd   &  93.28   &   22.09  \\  \hline                        
\end{tabular}
\end{table}

\begin{figure}[t]
    \centering
    \includegraphics[width=0.9\linewidth]{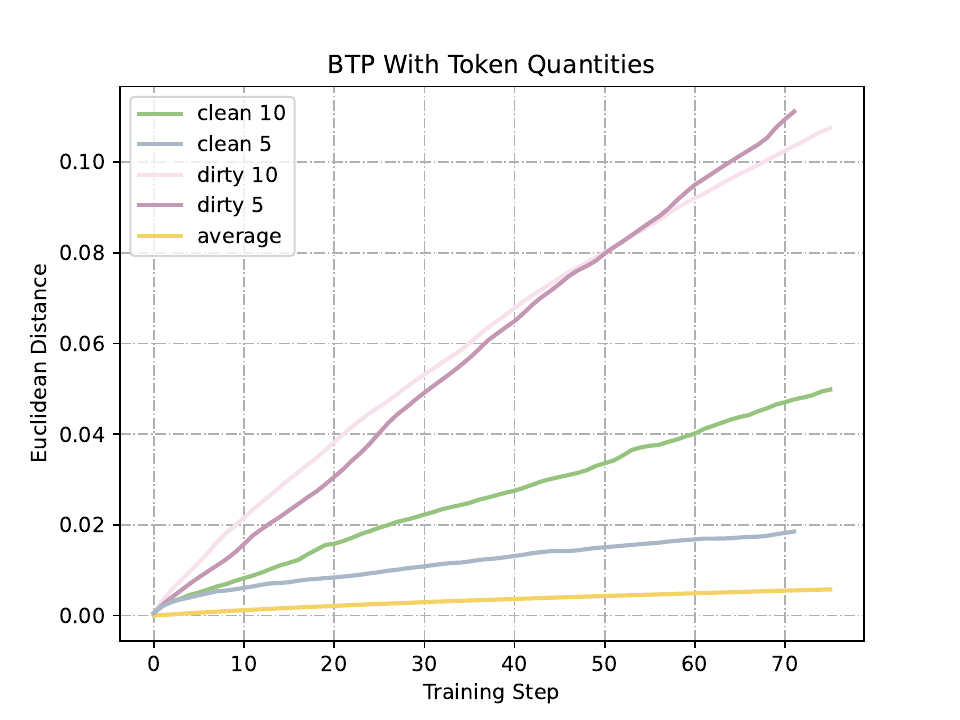}
    
    \caption{Token quantities influence the results. "clean" refers to not modifying the labels after insertion, "10" represents an insertion ratio of 10\%, and "average" indicates the mean of the changes in the word embedding layer.}
    \label{qualities}
\end{figure}

\subsection{Defense Results against Adaptive Attacks}
In this section, we consider the countermeasures an attacker might take when aware that the defender is using BTU. The core of BTU is to capture and purify backdoor tokens based on the simplicity of backdoor tasks. However, when backdoor tasks become more complex, backdoor tokens may evade detection, leading to potential defense failure. Therefore, adaptive attacks could be executed by narrowing the learning difficulty gap between backdoor and clean tasks. We will explore these potential adaptive attacks in the following discussion. \\

\noindent\textbf{Low Poison Ratio} In fact, a low poison ratio is more reflective of real-world scenarios. However, we found that most existing defense methods perform poorly against low poison ratio attacks. At the same time, a low poison ratio makes it more challenging for the model to learn the backdoor task. So we employ an experiment to test BUT's performance under low poison ratio backdoor attack. We use the lowest possible poison ratio (0.7\%) to perform Add-Sent attack on the SST-2 dataset, achieving an ASR of over 90\%. Then, we conducted training defense methods including RAP, Strip, BKI, ONION, CUBE, MF and BTU to evaluate their performances under low poison ratio backdoor attacks. As shown in Table~\ref{adaptive}, All defense methods failed to defend against low poison ratio backdoor attacks, except for BTU. This phenomenon shows that BTP exhibits a statistically significant change compared to CTP, even at a low poison ratio. This can be observed in Table~\ref{qualities}. It demonstrates that BTU possesses exceptionally strong backdoor defense capabilities. 

\noindent\textbf{Trigger Complexity} 
To increase trigger complexity, we adopt a method from the SOS~\cite{SOS} framework to perform adversarial training on a subset of triggers, thereby extending their length and complexity. This compels the model to learn the entire trigger sequence, heightening the learning challenge. As shown in Table~\ref{adaptive}, BTU effectively counters these enhanced backdoor attacks by identifying and neutralizing key trigger components during its exposure phase. These results demonstrate that BTU maintains strong defensive efficacy under varied and intensified backdoor conditions, effectively mitigating threats while adapting to increased task complexities. 
\subsection{Ablation Study}
\noindent\textbf{Token Quantities}
To evaluate whether the number of trigger tokens significantly affects BTP changes, we conducted an experiment with two models: one with random trigger insertions without label changes (clean) and another implementing a backdoor attack with both trigger insertions and label targeting (dirty). Each model incorporated triggers into 10\% of the dataset. We compared the changes in BTP after training the two models to assess the impact of token quantity. As shown in Fig.~\ref{qualities}, in backdoor attacks, BTP changes remain nearly unaffected by the number of tokens and are significantly higher than CTP with the same token count. These results demonstrate that our method robustly defends against backdoor attacks across varying poisoning rates.


\begin{table}[t]
\caption{Defense results of BTU against adaptive attack with low poison ratios and  complex triggers.}
\centering
\begin{tabular}{lrl}
\hline \textbf{Method} & \textbf{ASR} & \textbf{ACC} \\ \hline
None & 92.37 & 91.06 \\
BKI &95.29 & 90.74   \\
 RAP & 90.57 & 79.62  \\
STRIP & 95.72 & 90.72\\
ONION & 97.15 & 91.37\\
CUBE & 89.15 & 91.11 \\
MF & 38.27 & 91.05 \\
\textbf{BTU (ours)} & \textbf{7.36} & \textbf{90.39} \\
\hline
Trigger Complexity & 7.18& 90.71 \\
\hline
\end{tabular}
\label{adaptive}
\end{table}

\begin{table}[t]
\centering
\caption{Defense Results of BTU under different detection threshold, i.e. $\alpha$.}
\begin{tabular}{ccccc}
\hline
\multirow{2}{*}{\textbf{Method}} & \multicolumn{2}{c}{\textbf{Add-Sent}} & \multicolumn{2}{c}{\textbf{Synbkd}} \\
                        & ACC           & ASR          & ACC          & ASR         \\
\hline
None            & 91.03        & 100.00      & 90.75           & 91.52            \\ 
$\alpha=0.03$        & 91.05        & 5.91        & 90.56           & 37.91            \\ 
$\alpha=0.05$        & 90.96        & 4.84        & 90.17           & 24.77            \\ 
$\alpha=0.10$        & 87.56        & 4.87        & 89.73           & 15.96            \\ 
\hline
\end{tabular}
\label{cleantuneThres}
\end{table}
\begin{table}[t]
\caption{Compared with more token unlearning methods}
\centering
\begin{tabular}{ccccc}
\hline
\multirow{2}{*}{\textbf{Method}} & \multicolumn{2}{c}{\textbf{Add-Sent}} & \multicolumn{2}{c}{\textbf{Synbkd}} \\ 
                & \textbf{ACC} & \textbf{ASR} & \textbf{ACC} & \textbf{ASR} \\ 
\hline
None            & 91.03        & 100.00       & 90.48       & 86.89        \\ 
BTU-PN           & 88.86        & 13.82        & 90.01       & 26.00      \\ 
BTU-PR-1         & 90.87        & 99.88        & 90.59       & 43.56          \\ 
BTU-PR-2         & 90.24        & 4.91         & 89.70       & 29.71      \\ 
\textbf{BTU (ours)}         & \textbf{90.67 }       & \textbf{5.50}         & \textbf{90.47}      & \textbf{24.91}   \\ 
\hline
\end{tabular}
\label{unlearning}
\end{table}

\begin{table}[t]
\centering
\caption{Results for different anomaly detection rounds}
\label{tab:detection round}
\begin{tabular}{cccccc}
\hline
\multirow{2}{*}{\textbf{Anomaly Round}} & \multicolumn{2}{c}{\textbf{Add-Word}} & \multicolumn{2}{c}{\textbf{Synbkd}} \\
                       & \textbf{ACC} & \textbf{ASR} & \textbf{ACC} & \textbf{ASR} \\
\hline
None              & 91.06        & 100.0        & 90.72        & 90.48        \\
1                 & 90.55        & 17.21         & 90.63        & 29.89        \\
2                 & 90.60        & 7.81         & 90.60        & 37.49        \\
2+3               & 90.57        & 5.47         & 90.61        & 35.00        \\
1+2               & 90.35        & 7.70         & 90.57        & 27.70        \\
1+1+2+3           & 89.36        & 5.97         & 90.59        & 24.03        \\
\hline
\end{tabular}
\end{table}
\noindent\textbf{Detection Threshold $\alpha$}
To investigate the impact of the detection strength $\alpha$ on BTU, we conducted experiments on the SST-2 dataset with a 10\% poison ratio. Table~\ref{cleantuneThres} illustrates that adjusting the threshold value $\alpha$ plays a pivotal role in the efficacy of the BTU method. Increasing $\alpha$ enhances defense effectiveness but reduces model accuracy (ACC), while decreasing $\alpha$ preserves ACC but raises the attack success rate (ASR). Our findings suggest that setting $\alpha$ to 0.05 strikes an effective balance in defending against backdoor attacks across most scenarios. 

\noindent\textbf{Alternative Unlearning}
To further explore Token Unlearning, we tested three approaches:
\textbf{Parameter Replacement-1 (PR-1):} Following the insights from ~\cite{finemixing}, we replaced the BTP of the backdoored model with those from a pre-trained language model;
\textbf{Parameter Noise (PN):} Gaussian noise was added to the BTP to disrupt their backdoor characteristics; 
\textbf{Parameter Replacement-2 (PR-2):} We replaced the BTP of the backdoored model with those of the padding tokens. To further disrupt the BTP, we clipped dimensions in the BTP that showed significant changes, setting the clipping threshold to the mean change value of the CTP. We conducted experiments on the SST-2 dataset using the add-sent and synbkd attack methods with a 10\% poisoning rate. The results, detailed in Table~\ref{unlearning}, show that BTU outperforms other strategies.\\
\noindent\textbf{Anomaly Detection Rounds}
To assess the importance of each detection round, we conducted an ablation study on the SST-2 dataset, with results shown in Table \ref{tab:detection round}. We observe that the first detection round is more effective at mitigating complex backdoors, while the second and third rounds are better suited for countering simple backdoors. Increasing the number of detection rounds can reduce the success rate of backdoor attacks, though it may slightly impact accuracy. Our findings indicate that three detection rounds offer the optimal balance between maintaining accuracy and ensuring defense effectiveness. Additional ablation experiments are provided in the appendix.

\section{Conclusion}
In this work, we identified two key properties in the context of NLP backdoor learning: 1) the distinctive differences in the embedding values of backdoor tokens and clean tokens when only the word embedding layer is trained, and 2) the success of backdoor activation is highly related to the backdoor token parameters. Based on these observations, we propose a novel anti-backdoor learning method \textit{Backdoor Trigger Unlearning (BTU)}, which proactively exposes aberrant embedding parameters of backdoor tokens and mitigates backdoor behaviors during the training process. Extensive experimental results demonstrate that BTU can effectively defend against currently known backdoor attacks with minimal impact on the performance of clean tasks.

\noindent\textbf{Future Work} While BTU effectively defends against four different backdoor attacks and outperforms nine other defense methods, we cannot guarantee its effectiveness against more advanced future attacks. Further exploration is needed to provide theoretical guarantees for BTU's underlying mechanisms. Additionally, our current findings and defense results are based on evaluations with pre-trained language models, so it remains an open question whether BTU is effective for more advanced large language models.

\section{Acknowledgments}
This work was supported by the China National Science Foundation under Grant/Award Number 62072356 and the 111 Center (B16037). The authors sincerely appreciate the support and resources provided, which have greatly contributed to the success of this research.
\bibliography{aaai25}
\newpage
\clearpage

\section{Appendix}

\subsection{Experiment Details}

\textbf{Model and Dataset} The experiments were conducted on the BERT-BASE-UNCASED \cite{BERT} pre-trained model. The datasets used in this work are summarized in Table \ref{dataset}.\\
\textbf{Attack Setup} We adopted four representative backdoor attacks in the field of NLP: Add-Word \cite{badnets}, Add-Sent \cite{Add-Sent}, Stylebkd \cite{Stylebkd}, and Synbkd \cite{Synbkd}. These attacks utilize various trigger patterns, including word-level, sentence-level, and style-based triggers. We reproduced these attacks using the available open-source code \footnote{https://github.com/thunlp/OpenBackdoor} from OpenBackdoor. A summary of all the attacks evaluated in our work is provided in Table \ref{attack}.\\
\textbf{Defense Setup} We reproduced six mainstream defense methods: BKI\cite{BKI}, RAP\cite{RAP}, ONION\cite{ONION}, CUBE\cite{Openbackdoor}, MF\cite{MF}, and STRIP\cite{Strip}, using the open-source code from OpenBackdoor. These defenses cover a broad range of strategies, including backdoor sample detection, backdoor model detection, and anti-backdoor learning. The implementation details for these backdoor defense methods are provided below:
\begin{table*}[t]
\centering
\caption{Attack details}
\begin{tabular}{cccccc}
\hline
\textbf{Attacker} & \textbf{Poison Rate}  & \textbf{Epochs} & \textbf{BS} & \textbf{LR} & \textbf{Examples} \\ \hline
AddWord    & 0.1 &  5 & 32 & 2e-5  &final verdict \textbf{cf} : you 've seen it all before .\\
AddSent   &  0.1   & 5 & 32 & 2e-5 &final verdict : you 've seen it all before \textbf{I watched a 3D movie}.\\
Synbkd    & 0.1  & 5 & 32 & 2e-5 &\textbf{when you see it , you saw it before .}\\ 
Stylebkd  &  0.1  & 5 & 32 & 2e-5 &\textbf{Final verdict : have you seen it all before...}\\ \hline
\end{tabular}
\label{attack}
\end{table*}

\begin{itemize}
\item \textbf{BKI} ~\cite{BKI} detects backdoor words during the inference stage, this method identifies output logits and marks those with significant contributions as potential trigger words. We used the optimal hyperparameter settings recommended in the original paper, with $p$ set to 5 and $\alpha$ set to 0.1.

\item \textbf{RAP} ~\cite{RAP} detects backdoor samples during the inference stage by leveraging the robustness gap between clean and backdoor samples. We followed the optimal hyperparameter settings from the original paper, using 'tq' as the trigger and setting a probing range of $[-0.3, -0.1]$ to achieve optimal defense performance.

\item \textbf{ONION} ~\cite{ONION} detects backdoor words during both inference and training stages by evaluating the perplexity increase caused by trigger words. We adopted the optimal settings, with the threshold set to 0.

\item \textbf{CUBE} ~\cite{Openbackdoor} removes backdoor samples through a three-step process: mapping data into an embedding space, applying dimensionality reduction, and clustering to identify poisoned samples. We reproduced the experimental results using the code from OpenBackdoor.

\item \textbf{MF} ~\cite{MF} aims to reduce backdoor learning by moderating overfitting. We reproduced their results using the low-rank reparameterization method (LoRA).

\item \textbf{Strip} ~\cite{Strip} detects backdoor words during the inference stage by comparing the perturbation sensitivity between clean and poisoned text. We followed the optimal hyperparameter settings suggested in the original paper, repeating the process five times with a false rejection rate of 0.01.

\item \textbf{ST} ~\cite{ST} introduces a honeypot module to isolate backdoor functionality and employs a re-weighting mechanism to reduce the influence of backdoor samples on the main network. Since the code was not publicly available, we implemented a reproduction based on their paper.

\item \textbf{DPOE} ~\cite{DPOE} uses a trigger-only shallow model to capture spurious backdoor correlations, leaving the main model to learn residuals free from backdoor effects. We reproduced the experimental results using the publicly available code at \url{https://github.com/luka-group/DPoE}.

\item \textbf{TG} ~\cite{TG} leverages data partitioning and ensemble learning to limit the impact of backdoor trigger words. The experimental results were obtained using the publicly available code at \url{https://github.com/AI-secure/TextGuard}.

\item \textbf{BTU (Ours)} performs three rounds of anomaly token detection during the backdoor token detection stage. In each detection round, the model is trained for one epoch with a batch size of 32 and a learning rate of 2e-5. During the dimensional fine-grained unlearning stage, the model is further trained on clean data for three epochs with a learning rate of 5e-7. Specifically, for the SST-2 and OLID datasets, backdoor token detection is performed using the entire training dataset, while for the AG News dataset, it is conducted using a randomly selected 20\% subset of the training data.
\end{itemize}

\begin{table}[t]
\caption{Datasets used in our experiments.}
\centering
\begin{tabular}{cccccc}
\hline
\textbf{Dataset} & \textbf{Class} & \textbf{Avg Len} & \textbf{Train} & \textbf{Test} & \textbf{Dev} \\
\hline
SST-2            &2 &19.24 & 6920    & 1821        & 872         \\
OLID             &2 &25.40 & 11915   & 859         & 1323      \\
AG News          &4 &37.96 & 108000  & 7600        & 12000     \\
\hline
\end{tabular}
\label{dataset}
\end{table}

\begin{figure}[t]
    \centering
    \includegraphics[width=1\linewidth]{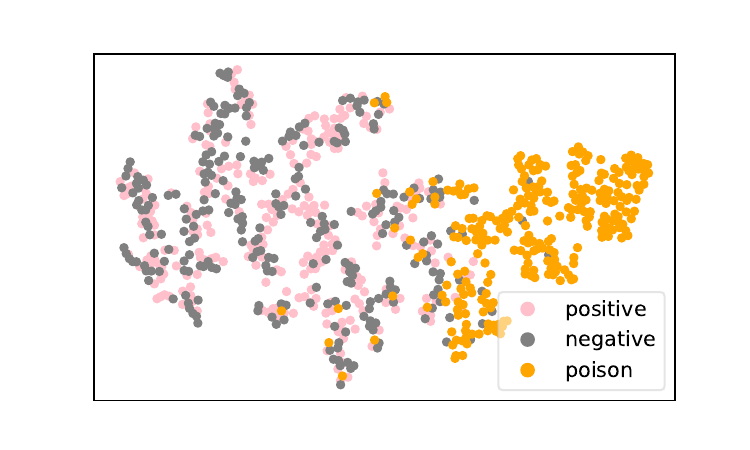}
    \caption{T-SNE visualization of CLS embedding parameters on a backdoored BERT. We trained the backdoored BERT model on the word embedding layers and plotted the output distribution between clean and poisoned samples. Note that 'positive' refers to examples with a ground-truth label of 1, and 'negative' refers to examples with a ground-truth label of 0. It is observed that most poisoned examples (marked in yellow) are clustered separately from the others.}
    \label{tsne}
\end{figure}

\subsection{Visualization of Output Distribution on Word Embedding Parameters}

Figure~\ref{tsne} presents the T-SNE visualization of the backdoored CLS embedding parameters. We applied the Add-Sent attack on the SST-2 dataset and extracted the CLS embeddings for both clean and poisoned samples. As shown in Figure~\ref{tsne}, while there is no clear separation among the clean samples, a distinct boundary exists between the poisoned and clean samples. This suggests that training solely on the word embedding layer effectively captures the backdoor information in our BTU detection.

\subsection{Results of BTU against Multi-trigger Attacks}

Recently, Multi-Trigger Backdoor Attacks (MTBAs)~\cite{multiattack} have emerged as a new threat, where multiple adversaries use different types of triggers to poison the same dataset, posing significant challenges to existing defenses. MTBA introduces three novel paradigms: single trigger to a single target (One2One), multiple triggers to a single target (All2One), and multiple triggers to multiple targets (All2All). 

To evaluate the effectiveness of our BTU defense, we evaluate BTU against MTBAs across all three attack modes on the SST-2 dataset. Table 8 reports the defense results. It shows that BTU is highly effective in mitigating MTBAs, reducing the ASR from 100.00 to an average of 6.69 on Add-Word and from 100.00 to an average of 15.57 on Add-Sent, while the accuracy (ACC) remains nearly unchanged. This result verifies that the generalizability and robustness of BTU defend against more advanced backdoor attacks.

\begin{table}[ht]
\caption{BTU defense against MTBAs}
\centering
\begin{tabular}{ccccc}
\hline
\multirow{2}{*}{\textbf{Method}} & \multicolumn{2}{c}{\textbf{Add-Word}} & \multicolumn{2}{c}{\textbf{Add-Sent}} \\
                        & \textbf{ACC}           & \textbf{ASR}          & \textbf{ACC}           & \textbf{ASR}\\
\hline
W/O Defense             & 91.05         & 100.00      & 91.05         & 100.00        \\
All2One (3)             & 90.76         & 7.99        & 90.59         & 13.93        \\
All2One (5)             & 90.81         & 5.75        & 90.47         & 25.08        \\
All2All (2)           & 90.39         & 7.14        & 90.70         & 7.71       \\
\hline
\end{tabular}
\label{attackmode}
\end{table}

\subsection{Analysis of Backdoor Parameters Detection}

Here, we provide a theoretical analysis for BTU defense. Formally, backdoor learning can be viewed as a dual-optimization with a clean task and a backdoor task. The optimization of backdoor attack on embedding parameters can be re-defined as follows:

\[
\varepsilon^{c}_{\text{new}} = \varepsilon^{c} + \sum\limits_{i=1}^{\text{step}} \partial_i^c,
\]

\[
\varepsilon^{b}_{\text{new}} = \varepsilon^b + \sum\limits_{i=1}^{\text{step}} \partial_i^c + \sum\limits_{i=1}^{\text{step'}} \partial_i^b,
\]

\noindent where $\partial$ denotes the gradient at each step, and $\varepsilon^{c}_{\text{new}}$ and $\varepsilon^{b}_{\text{new}}$ represent the clean token embedding parameters (CTP) and backdoor token embedding parameters (BTP) respectively. 

We aim to distinguish between BTP and CTP by analyzing the change in parameters. To measure the change in word embedding parameters before and after training, we use Euclidean distance as the metric. The specific calculation method is as follows:

\[
d_c = d(\varepsilon^{c}_{\text{new}}, \varepsilon^{c}) \approx ||\sum\limits_{i=1}^{\text{step}} \partial_i^c||_2
\]

\[
d_b = d(\varepsilon^{b}_{\text{new}} , \varepsilon^b) \approx ||\sum\limits_{i=1}^{\text{step}} \partial_i^c + \sum\limits_{i=1}^{\text{step'}} \partial_i^b||_2
\]

where $d$ represents the Euclidean distance. As the BTP experiences a more significant change compared to CTP, it causes effective backdoor parameter detection in our BTU defense.
\begin{algorithm}[!t]
\caption{BTU: Backdoor Token Unlearning}
\label{alg:expose}
\textbf{Stage 1}: Backdoor Token Detection\\
\textbf{Input}: Dataset: $\mathcal{D}$, PLM: $M=(\varepsilon,..., cls)$, threshold: $\alpha$\\
\textbf{Output}: Backdoor token set $T$
\begin{algorithmic}[1] 
\STATE Let $T=T_1=T_2=T_3=\phi$, $\varepsilon$ denotes embedding.
\STATE Train $\varepsilon$ of $M$ by $\mathcal{D}$ $\rightarrow \varepsilon'$.
\STATE Get $T_1$ by Eq. \ref{t1}
\STATE Sort $s_i'$ in set $T_1$ in descending order. 
\STATE Extract the top $\alpha$\% of tokens as $T'=\{t_i\}_{i=1}^{|\mathcal{V}|\times \alpha}$
\STATE Train $\varepsilon$ of $M^*=(\varepsilon, c)$ by $\mathcal{D}$ $\rightarrow \varepsilon''$.
\STATE Get $T_2$ by Eq. \ref{t2}
\STATE Sort $s_i''$ in set $T_2$ in descending order. 
\STATE Extract the top $\alpha$\% of tokens as $T''=\{t_i\}_{i=1}^{|\mathcal{V}|\times \alpha}$
\STATE Train $\varepsilon$ of $M^*$ by $\mathcal{D}/T''$ $\rightarrow \varepsilon'''$.
\STATE Get $T_3$ by Eq. \ref{t3}
\STATE Sort $s_i'''$ in set $T_2$ in descending order. 
\STATE Extract the top $\alpha$\% of tokens as $T'''=\{t_i\}_{i=1}^{|\mathcal{V}|\times \alpha}$
\STATE \textbf{return}  $T=T'\cup T''\cup T'''$
\end{algorithmic}
\textbf{Stage 2}: Dimensional Fine-gained Unlearning\\
\textbf{Input}: $T$, padding token $p$, backdoored model $M^b$, clean dataset $\mathcal{D}'$\\
\textbf{Output}: clean model $M^c$
\begin{algorithmic}[1]
\STATE Get $\varepsilon^b$ from $M^b$, compute $\bar{d}$ by Eq. \ref{eq6}.
\STATE Let $\varepsilon^c=\varepsilon^b$.
\FOR{$t$ in $T$}
\STATE $\varepsilon_i^c(t) =
\begin{cases} 
\varepsilon_i'(t), & \text{if } |\varepsilon_i'(t) - \varepsilon_i(t)| < \bar{d}; \\
\varepsilon_i'(p), & \text{if } |\varepsilon_i'(t) - \varepsilon_i(t)| \geq \bar{d};
\end{cases}$
\ENDFOR
\STATE Replace $M^b$ embedding with $\varepsilon^c$ to get $M^c$
\STATE Fine-tune the $M^c$ by $\mathcal{D}'$
\STATE \textbf{return}  $M^c$
\end{algorithmic}
\label{algorithnm}
\end{algorithm}
\begin{table}[ht]
\centering
\caption{Anomaly Detection in Other layers}
\label{tab:layer_performance}
\begin{tabular}{ccccc}
\hline
\textbf{Layer} & \multicolumn{2}{c}{\textbf{Add-Word}} & \multicolumn{2}{c}{\textbf{SynBkd}} \\
               & \textbf{ACC} & \textbf{ASR} & \textbf{ACC} & \textbf{ASR} \\
\hline
None           & 91.05        & 100.00       & 90.72        & 90.48        \\
1              & 61.56        & 22.69        & 81.08        & 76.43        \\
2              & 58.25        & 87.64        & 52.77        & 49.20        \\
3              & 50.85        & 81.98        & 63.42        & 62.96        \\
4              & 58.34        & 91.21        & 60.21        & 55.73        \\
5              & 68.20        & 32.51        & 50.11        & 53.32        \\
\hline
\end{tabular}
\end{table}
\begin{table}[ht]
\centering
\caption{Performance on more datasets}
\label{tab:dataset_comparison}
\begin{tabular}{cccccc}
\hline
\multirow{2}{*}{\textbf{Dataset}} & \multirow{2}{*}{\textbf{Defense}} & \multicolumn{2}{c}{\textbf{Add-Sent}} & \multicolumn{2}{c}{\textbf{Synbkd}} \\

                 &                  & \textbf{ACC} & \textbf{ASR} & \textbf{ACC} & \textbf{ASR} \\
\hline
\multirow{2}{*}{IMDB} 
             & None             & 90.83        & 100.00       & 90.30        & 97.15        \\
             & BTU       & \textbf{90.64}        & \textbf{1.07}         & \textbf{90.04}        & \textbf{25.18}        \\
\hline
\multirow{2}{*}{Offens}
       & None             & 81.30        & 100.00       & 81.29        & 95.66        \\
       & BTU       & \textbf{80.86}        & \textbf{8.31}         & \textbf{81.73}        & \textbf{11.28}        \\
\hline
\end{tabular}
\end{table}

\subsection{Results of More Layers}
To investigate whether parameter changes in other model layers are more closely linked to backdoor behavior, we conducted experiments on the BERT model using the SST-2 dataset, focusing on Add-Word and SynBkd backdoor attacks. The analysis targeted the five bottom layers of the model, as previous research suggests that the lower layers are more likely to capture backdoor features [1]. During the experiments, we zeroed in parameters in each layer that exhibited changes that exceeded the mean parameter change for that layer after backdoor training. We then evaluated both the attack success rate (ASR) and the clean accuracy (ACC). The results are shown in Table \ref{tab:layer_performance}. The findings reveal that, aside from the word embedding layer, parameter changes in other layers are not strongly correlated with the presence of a backdoor.

\subsection{Results of More Datasets}
We expanded our experiments to include two additional datasets: Offens \cite{OffensEval} and the Internet Movie Database (IMDB) \cite{IMDB} to further evaluate the effectiveness of our method. The results presented in Table \ref{tab:dataset_comparison} demonstrate that BTU performs well in multiple datasets.

\end{document}